\documentclass[11pt]{article}
\usepackage{epsfig}
\usepackage{a4}
\usepackage{amssymb}
\usepackage{amsfonts}

\begin{document}

\def\Mpeppo{p \rightarrow e^+ \pi^0}
\def\Mpmppo{p \rightarrow \mu^+ \pi^0}
\def\Mpepeta{p \rightarrow e^+ \eta}
\def\Mpmpeta{p \rightarrow \mu^+ \eta}
\def\Mnnueta{n \rightarrow \bar{\nu} \eta}
\def\Mpnukp{p \rightarrow \bar{\nu} K^+}
\def\Mpmpko{p \rightarrow \mu^+ K^0}
\def\Mpepko{p \rightarrow e^+ K^0}
\def\Mnnuko{n \rightarrow \bar{\nu} K^0}
\def\peppo{\(\Mpeppo{}\)}
\def\pmppo{\(\Mpmppo{}\)}
\def\pepeta{\(\Mpepeta{}\)}
\def\pmpeta{\(\Mpmpeta{}\)}
\def\nnueta{\(\Mnnueta{}\)}
\def\pnukp{\(\Mpnukp{}\)}
\def\pmpko{\(\Mpmpko{}\)}
\def\pepko{\(\Mpepko{}\)}
\def\nnuko{\(\Mnnuko{}\)}
\def\Mnumunog{(\not\!\gamma)K^+ \rightarrow (\not\!\gamma)\mu^+ \nu_{\mu}}
\def\Mpnukpnumunog{\Mpnukp ;\; \Mnumunog}
\def\Mnumug{(\gamma)K^+ \rightarrow (\gamma)\mu^+ \nu_{\mu}}
\def\Mpnukpnumug{\Mpnukp ;\; \Mnumug}
\def\Mnumu{K^+ \rightarrow \mu^+ \nu_{\mu}}
\def\Mpnukpnumu{\Mpnukp ;\; \Mnumu}
\def\Mpippio{K^+ \rightarrow \pi^+ \pi^0}
\def\Mpnukppippio{\Mpnukp ;\; \Mpippio}
\def\pnumunog{\(\Mnumunog\)}
\def\pnumug{\(\Mnumug\)}
\def\numu{\(\Mnumu\)}
\def\pippio{\(\Mpippio\)}
\def\pnukpnumunog{\(\Mpnukpnumunog\)}
\def\pnukpnumug{\(\Mpnukpnumug\)}
\def\pnukpnumu{\(\Mpnukpnumu\)}
\def\pnukppippio{\(\Mpnukppippio\)}
\newcommand{\cllimitny}[3] {\(\tau/B_{#1} > #2 \times 10^{#3}\)}
\newcommand{\cllimit}[3] {\cllimitny{#1}{#2}{#3} years (90\% CL)}

\begin{flushright}
\today
\end{flushright}
\vspace*{1cm}
\begin{center}
{\Large{\bf Very massive underground detectors for proton decay searches\footnote{Based on an invited
talk at the XI International Conference on Calorimetry in High Energy Physics - CALOR2004,
Perugia, Italy, March 2004.}}}\\
\vspace{0.5cm}
{\large A. Rubbia}

Institut f\"{u}r Teilchenphysik, ETHZ, CH-8093 Z\"{u}rich,
Switzerland
\end{center}
\begin{abstract}
\noindent
Massive underground detectors can be considered as sort of observatories
for rare physics phenomena like astrophysical neutrino detection and nucleon decay
searches. We briefly overview the past tracking calorimeters developed for nucleon
decay searches. We then discuss the two technologies which are discussed today for
potential future applications: Water Cerenkov ring imaging (WC) and liquid Argon Time Projection
Chamber (LAr TPC). We present a conceptual design for a 100~kton liquid Argon TPC.
We illustrate the physics performance of Water Cerenkov and liquid Argon TPC
detectors for the $p\rightarrow e^+\pi^0$ and $p\rightarrow \nu K^+$
proton decay searches. We briefly compare the physics reach of the two techniques.
We conclude by stressing the complementarity of the two approaches, noting however that, given the
foreseeable timescale for these next generation experiments, the new challenging technique of the
LAr TPC might offer more discovery potentials.
\end{abstract}

\section{Introduction}

\subsection{The physics of massive underground detectors}
Most massive underground detectors were designed by optimizing their performance
for the search of nucleon decays. In this short review, we will discuss these detectors
under this point of view, however, massive underground
detectors have a much larger physics program, for example with the
observation and study of astrophysical (solar, atmospheric, and supernova neutrinos)
and artificial beam neutrinos. Such a comprehensive physics program,
possibly with non-accelerator and accelerator-based components, makes
massive underground detectors ``general purpose'' facilities, sort of observatories
for rare physics phenomena.

Nucleon decay studies are presently in a ``second generation'' phase, after
the enthusiasm following the development of the first GUT theories. The
minimal version of SU(5)\cite{gg}, predicting the decay 
$p\to e^+\pi^0$
with a lifetime of about $10^{31}$ years, has been ruled out
by the experimental limits ($\tau > 2.9\times 10^{33}$ years), and also fails
to predict the correct value of $\sin^2\theta_W$.
Alternative models have been proposed, for instance SUSY GUTs, that 
predict values of $\sin^2\theta_W$ closer to the measured ones, and the 
presence of an intermediate mass scale of $O(M_W)$, that seems to provide a
better unification of the coupling constants, at higher values with respect
to minimal SU(5)\cite{amal}.
The higher unification mass pushes up the proton lifetime in the 
$p\to e^+\pi^0$ channel, which has predicted lifetimes of $10^{36\pm1}$ years,
compatible with the experimental limits.
However, in this scenario, other decay channels open up, where supersymmetric
intermediate states are involved. In particular, s-quark production is
favoured, and final states involving kaons (like $p\to K^+\nu$) are present.

\subsection{Detector optimization}
Nucleon decay signals are characterized by (a) their topology (b) their kinematics.
By topology, we mean the necessary presence of a lepton (an electron, a muon or a
neutrino) in the final state, in general, few particles in the end products (for example,
two body decays are believed to be favored), and obviously no other 
energetic nucleon in the final state. The exact kinematics of the event depends
on the type of target. For free protons (target with hydrogen), the total
momentum of the event should be compatible with zero, while for nucleon
decays occurring in nuclear targets, we expect a smearing from Fermi motion and
also other nuclear effects (rescattering, absorption, etc.).
The total energy of the event 
should be equal to the nucleon mass, which means
in the GeV range.

As far as detector optimizations are concerned, we stress three points needed to 
reach stringent sensitivities on nucleon lifetime with low backgrounds:
(1) the necessity of fine tracking (2) the necessity of excellent resolution
(3) the necessity of large masses. We discuss these points in the following.

The fine tracking (hence high spatial resolution) is a fundamental tool for
the visualization of the event and for the particle identification. This is of extreme
importance to separate potential signal from background. Discovery experiments
should by definition be low background, in order to make
possible claims based on few events. Ideally, one would like to have
3 dimensional tracking and vertex reconstruction. For particle identification,
stopping power measurement ($dE/dx$) is important to separate low
momentum particles. Integration of $dE/dx$ yields the kinetic energy. 
The separation of electrons, muons,
protons, pions and kaons in the momentum ranges between $\simeq 10$ and $\simeq 100$~MeV/c
is wanted.
Good particle identification
is specially required for $\mu/\pi$ and $e/\pi^0$ separation.

\begin{table}[tbh]
\caption{Massive underground experiments for nucleon decay searches}
{
\begin{tabular}{@{}lcllc@{}}
\hline
{} &{} &{} &{} &{}\\[-1.5ex]
Detector & Date & Type & Location & Typ. mass \\[1ex]
\hline
{} &{} &{} &{} &{}\\[-1.5ex]
NUSEX & 1982 & Tracking calorimeter & Mont-Blanc, France  & 0.15 kton \\[1ex]
Fr\'ejus & 1985 & Tracking calorimeter & Fr\'ejus, France  & 0.9 kton \\[1ex]
Soudan & 1989 & Tracking calorimeter & Minnesota, USA  & 0.96 kton \\[1ex]
\hline
Kamiokande & 1983 & Water Cerenkov & Gifu, Japan & 0.88 kton \\[1ex]
IMB-3 & 1986 & Water Cerenkov & Ohio, Japan & 3.3 kton \\[1ex]
Super-Kamiokande & 1996 & Water Cerenkov & Gifu, Japan & 22.5 kton \\[1ex]
\hline
ICARUS & 2006(?) & Liquid Argon & Gran Sasso, Italy & 0.6$\rightarrow$ 3 kton \\[1ex]
\hline
\end{tabular}\label{tab:nucdecexp} }
\end{table}

Precise calorimetric information in the GeV region requires high sampling rate
and containment of the particles and showers (e, $\pi^0$, ...). The shower direction
is also needed for kinematical reconstruction. Calorimetric performance
has an impact on the granularity of the detector.

\begin{figure}[bt]
\centerline{\epsfxsize=6.1in\epsfbox{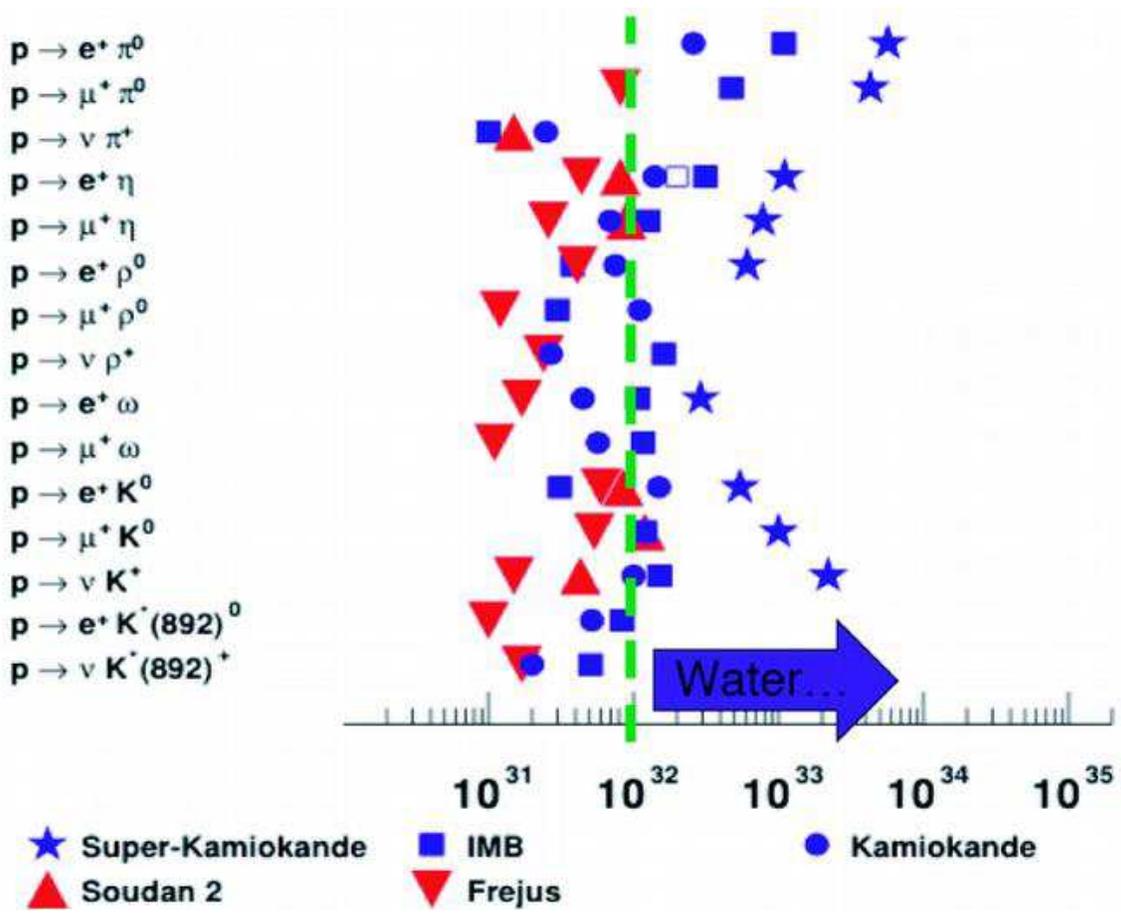}}   
\caption{Illustration of current experimental limits on nucleon lifetime. \label{fig:curlimits}}
\end{figure}

Finally, the mass scale of the detectors is given by the nucleon lifetime scale.
Since there is about $6\times 10^{32}$ nucleons per kton of mass, the proton
lifetime limit in case of no backgrounds (and no signal!)  is simply given by:
\begin{equation} 
\tau_p/Br\ (90\% C.L.) > 10^{32} years \times M(kton) \times T(yr) \times \epsilon
\end{equation}
where $M$ is the detector mass in kton, $T$ is the exposure in years
and $\epsilon$ is the signal detection efficiency after cuts, which depends
on the considered decay channel. 
If background (mainly coming from atmospheric neutrinos when the detector is located
deep underground) is present, the
sensitivity increases only proportionally to $\sqrt{M}$. 

Given the variety
of predicted decay modes, the ideal detectors should be as versatile as
possible, very good in background rejection, and at the same time have
the largest possible mass.
The relevant factor
is in fact $Mass\times \epsilon$, hence, large masses must be coupled 
to fine tracking and excellent calorimetry, to suppress
backgrounds with a good signal detector efficiency.

\section{Nucleon decay experiments}
A list of massive underground experiments for nucleon decay searches is given in Table~\ref{tab:nucdecexp}.
Their typical masses are given in the last column of the table.
We note that NUSEX, Fr\'ejus, Soudan, Kamiokande and IMB stopped data-taking and currently
only Superkamiokande is active.
ICARUS is under construction and is the only planned new experiment. Other bigger detectors
are being discussed with timescales
likely beyond the year 2015 and will be mentioned later. 

A summary of the current limits on nucleon
decays is illustrated in Figure~\ref{fig:curlimits}. We note a clear separation
line at the level of $\simeq 10^{32}$ years. Above this value, Water Cerenkov (in fact, Superkamiokande)
results dominate. This shows that the factor for improving
``limits'' in nucleon decay has been mass. The fine granularity calorimeter, which provided a priori
a better information on events, did not yield limits beyond $\simeq 10^{32}$ years because
of their limited mass. The best limits are currently in the range of $\simeq 10^{33}$ years.
As will be shown later, improving sensitivities beyond the currently achieved results will
require very large masses but also good detector performances. Indeed, the Water Cerenkov
technique is reaching the level at which background events start to become reducible only
at the cost of signal efficiency. In this regime, the gain in sensitivity will not be linear with mass
anymore. We will show that the liquid Argon TPC technique provides better performance than Water Cerenkov
detectors and can therefore solve the problem of backgrounds, so the question is how large a liquid Argon 
TPC can be? Before we discuss this point,
we summarize in the following sections (1) the basic features of the tracking calorimeters and (2)
the Water Cerenkov detectors and their performance for two proton decay channels.

\section{Tracking calorimeters}

\subsection{NUSEX experiment}
NUSEX\cite{Battistoni:1985na}
 was composed of iron plates interleaved with streamer tubes. Its average
density was 3.5~g/cm$^3$. The total mass was 150~tons. In total there were 134
horizontal iron plates, each plate has dimensions of 3.5~m $\times$ 3.5~m $\times$ 1 cm.
The streamer tubes were made of plastic (3.5 m long, $9\times 9$~mm$^2$ cross-section)
with digital readout. The tracking was provided by the planes of tubes equipped
with X \& Y pickup strips for two dimensional localization.
The total number of tubes was 42880. The gas mixture was $Ar-CO_2(15\%)-n pentane (1+2+1)$.
The resolution was 0.3~cm in the X-view, 0.37~cm in Y and the track separation was about
1.5~cm. A test module was exposed to electrons at the CERN PS with momenta
ranging from 0.150 to 2~GeV/c at varying angles ($0\rightarrow 40^o$). The energy
resolution was $\sigma/E \approx 0.13/\sqrt{E(GeV)}\oplus 0.2$. Muons and pions
momentum was estimated with range. Poor separation between them was 
possible. One particularity was the exposure of a test module to actual neutrinos
at CERN. A beam of 10~GeV protons was sent on a Be target which was configured
in a beam-dump mode. The decay length was 10~m and was followed by 12~m of
Fe/concrete shielding. The detector was exposed at two angles and the following
statistics was collected: 210 (0 deg) and 184 (45 deg) neutrino interactions were
identified. These were used directly to estimate the background in nucleon
decay searches. The nucleon decay 
results of NUSEX can be found in Ref.\cite{Battistoni:1983ka}. It is interesting
to note the existence of a candidate in the $p\rightarrow \mu^+K^0$ mode
which had an estimated background of 0.09.

\subsection{Fr\'ejus experiment}
The Fr\'ejus experiment\cite{Berger:1987ke} was a modular tracking calorimeter composed
of 912 layers of sandwich of 3~mm thick iron and $5\times 5$~mm$^2$ plastic
flash tubes coupled with  $15\times 15$~mm$^2$ Geiger tubes for triggering. The total mass was 900~tons
and the average density $1.95$~g/cm$^3$. The detector dimensions were
6~m (wide) $\times$ 12.3~m (deep) $\times$ 6~m (high). The absorber was iron. The readout
was done via the 6~m long flash tubes whose discharge was triggered by a signal
in the corresponding Geiger tubes. The flash tubes contained a Neon(70\%)-Helium(30\%) mixture
and were pulsed with 5~kV for a duration $>800$~ns with a fast rise-time ($<100$~ns)
not to act as a clearing field. The long time was needed to propagate the discharge down the 6~m long
tubes.
The tracking was obtained by the orientation
of the planes in various directions. In total, there were $10^6$~tubes. The spatial resolution
was about 5~mm.
The electromagnetic
resolution was $\sigma/E \approx 0.06/\sqrt{E(GeV)}\oplus 0.055$. A pizero mass distribution was
reconstructed and had a width of 16\%. Muons and pions
momentum was estimated with range. Poor separation between them was 
possible. Results on nucleon decay can be found in Ref.\cite{Berger:1991fa}. The
relatively good tracking performance of the system can be appreciated
in the list of final states searched by the Fr\'ejus experiment. 

\subsection{Soudan experiment}
The Soudan-II experiment\cite{Thron:1989cd} was characterized as a fine grain, finely segmented calorimeter.
The detector was composed of 224 calorimeter modules for a total mass of 960 tons.
Each module had a mass of 4.3~tons and a volume of 1~m (wide) $\times$ 1.11~m (deep)
$\times$ 2.7~m (high). The absorber was composed of corrugated iron sheets yielding
an average density of 1.6~g/cm$^3$. The readout was made possible by 1~m long drift
tubes with a diameter of 1.5~cm. Tracking was provided by the tubes readout which allowed
for three spatial coordinates and $dE/dx$ measurement. The total number of tubes
was $\approx 2\times 10^6$. The maximum drift in the tubes was 50~cm and the gas
mixture was $Ar-CO_2(15\%)$ yielding a drift velocity of 0.6~cm/$\mu s$ at 180~V/cm.
The resolutions were 0.38~cm in the xy-plane and 0.65~cm in the z-direction. Nucleon
decay results
from the Soudan-II experiment can be found in Ref.\cite{goodmanicrc}.

\section{Water Cerenkov detectors}
\subsection{The technique}
This technique was developed in the first large scale detectors
Kamiokande\cite{Koshiba:mw} and IMB\cite{Becker-Szendy:hr}.
The largest detector existing today is the Superkamiokande\cite{Fukuda:2002uc}
Water Cerenkov detector. It is composed of a tank of 50~kton of water (22.5~kton fiducial) which is
surrounded by 11146 20-inch phototubes immersed in the water. About 170~$\gamma$/cm are produced
by relativistic particles in water in the visible wavelength $350<\lambda<500$~nm. With 40\% PMT coverage
and a quantum efficiency of 20\%, this yields $\approx 14$ photoelectrons per cm or $\approx 7$~p.e. per MeV 
deposited.
Contrary to the fine tracking calorimeters described in the previous section,
water provides free protons in 20\% of the cases, which should yield events without Fermi smearing.
The performance of the Superkamiokande detector is good, regardless of its coarse imaging nature.
The vertex resolution is about 30~cm for 1-ring events and estimated to be
15~cm for $p\rightarrow e^+\pi^0$ events. The trigger threshold is set at 5~MeV photoelectron equivalent,
which means the triggering of nucleon decay is 100\% efficient. The energy resolution is
approximately 3--4\% for electrons and muons. Strikingly enough particle identification is made
possible by the features of the Cerenkov rings: fuzzyness signals electrons while sharp-edged rings
indicate muons. The particle separation is better than 99\% for muons and electrons. In case of
$p\rightarrow e^+\pi^0$  and $p\rightarrow \mu^+\pi^0$ it is greater than 97\%. One major drawback
is the high Cerenkov momentum threshold, namely 568~MeV/c for kaons and 1070~MeV/c for protons!
This implies for example that kaons produced in $p\rightarrow K^+\nu$ are invisible.

\begin{figure}[tb]
\centerline{\epsfxsize=5.1in\epsfbox{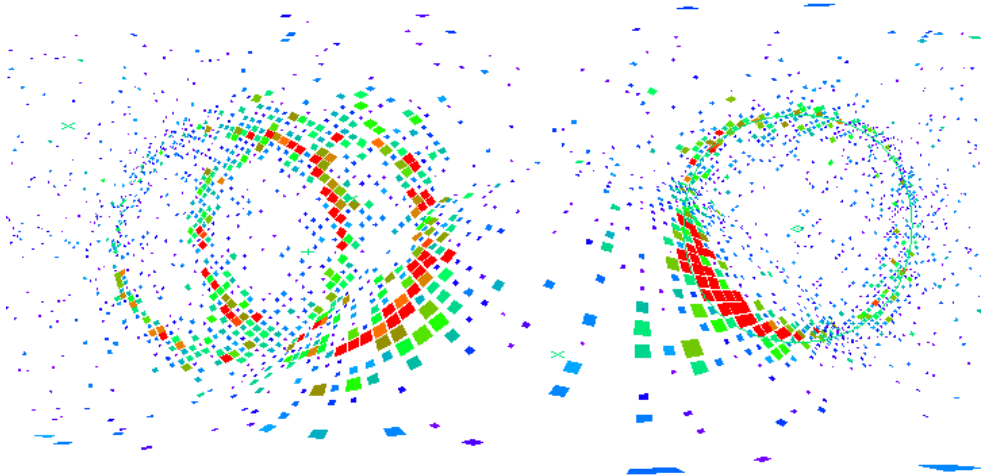}}   
\caption{MC event of $p\rightarrow e^+\pi^0$ in Superkamiokande\protect\cite{Viren:1999pk}. \label{fig:mcepi0}}
\end{figure}

\subsection{Results on proton decay}
In Water Cerenkov detectors, an ideal \peppo{} event might look like
Figure~\ref{fig:mcepi0}.  This event was generated with a detailed
\peppo{} event and detector Monte Carlo (MC) simulation\cite{Viren:1999pk}. In this
figure, the PMTs are plotted as a function of \(\cos\theta\) {\it vs.}
\(\phi\) as viewed from the event vertex and are represented by
squares, colored by amount of collected charge (red is more, blue is
less) and sized to show distance from the event vertex.  The fuzzy
outer edges of the rings indicate an electromagnetic showering type of
ring. 

\begin{figure}[htb]
  \centering
  \begin{minipage}[t]{2in}
    \epsfxsize=2in \epsfbox{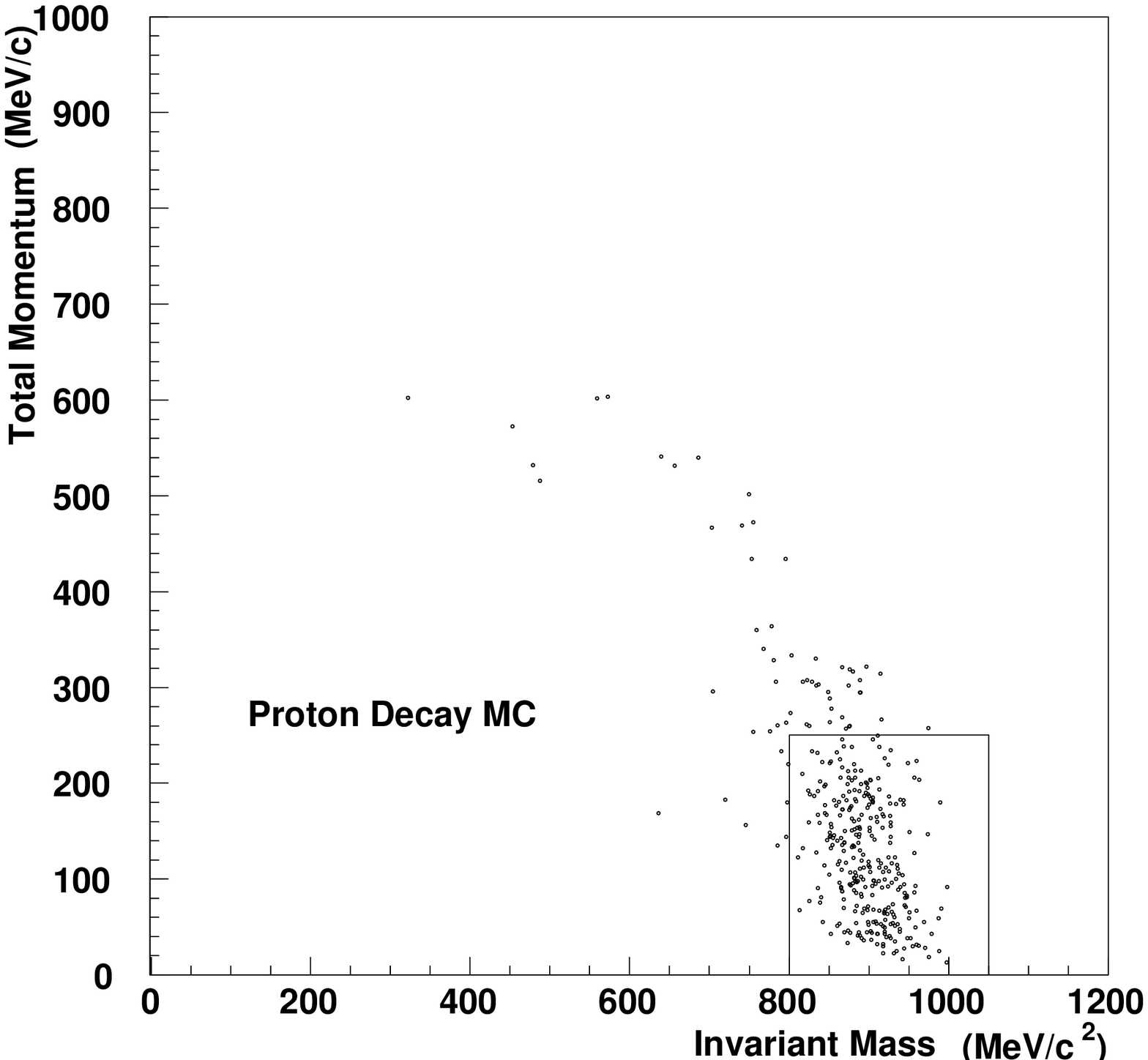}
  \end{minipage}
  \begin{minipage}[t]{2in}
    \epsfxsize=2in \epsfbox{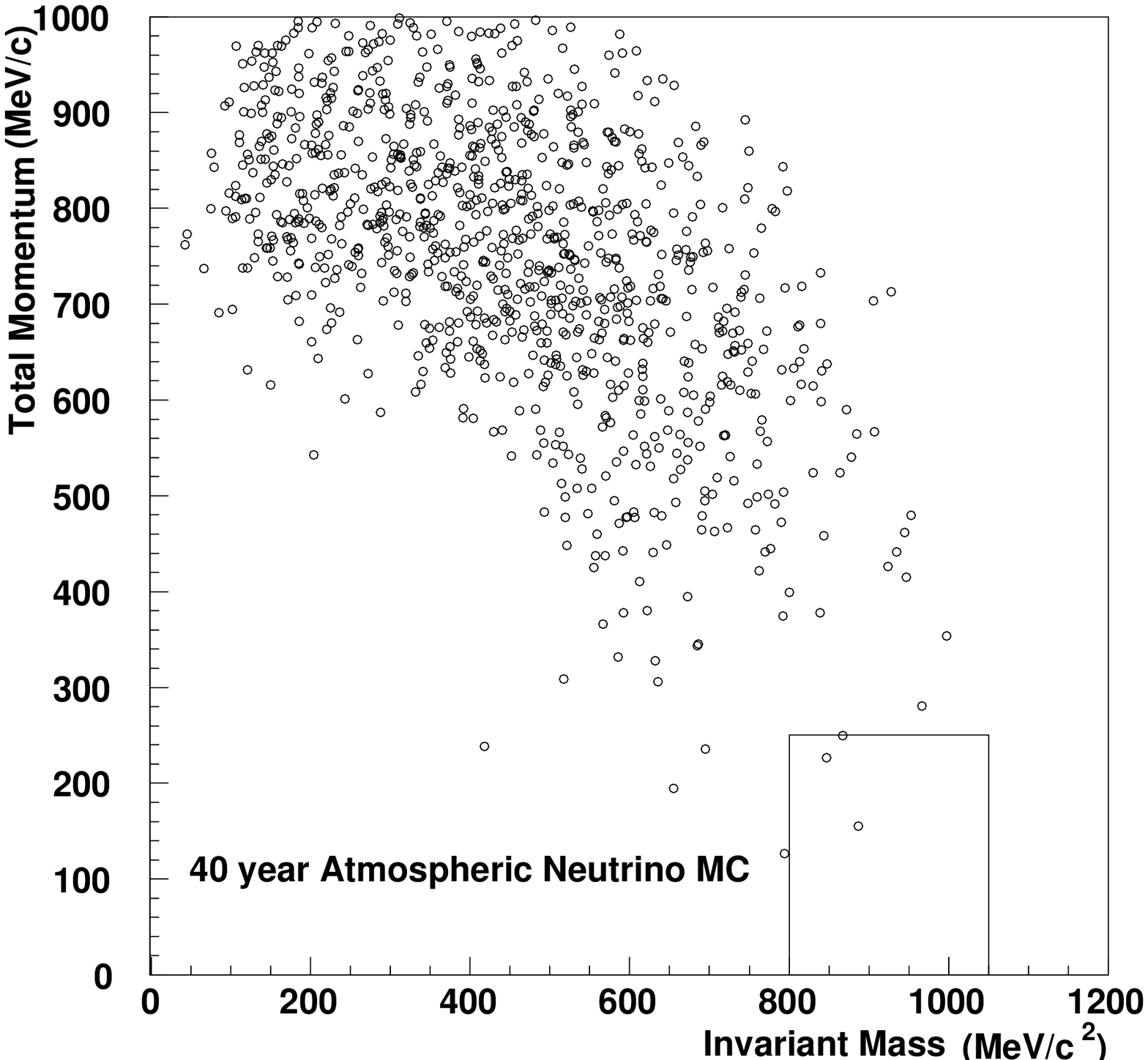}
  \end{minipage}
  \begin{minipage}[t]{2in}
    \epsfxsize=2in \epsfbox{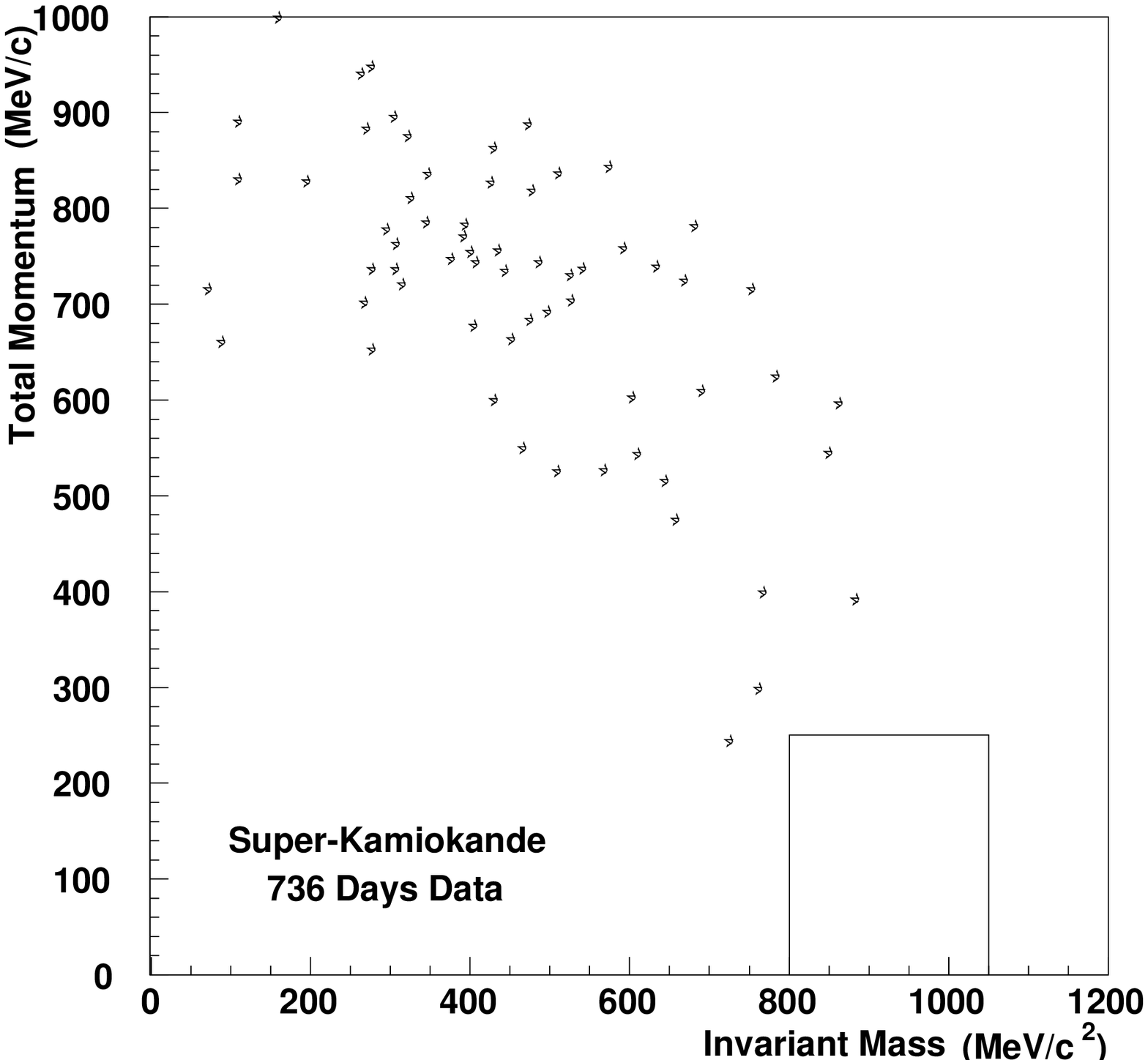}
  \end{minipage}
  \caption{\peppo{} mode.  Distributions of events in total reconstructed momentum {\it vs.} total invariant mass for (a) proton decay MC, (b) atmospheric neutrino MC, and (c) data.}
  \label{fig:peppo-massmom}
\end{figure}

Superkamiokande searches for the \peppo{} mode with selection criteria that are as follows\cite{Viren:1999pk}
(See Table~\ref{tab:1sk}): 
(A)~6000 \(< Q_{tot} <\) 9500 photoelectrons (PEs), 
(B)~2 or 3 e-like (showering type) rings, 
(C)~if 3 rings: 85 \(< M_{inv,\pi^0} <\) 185 MeV/c\(^2\), 
(D)~no decay electrons, 
(E)~800 MeV/c\(^2\) \(< M_{inv,tot} <\) 1050 MeV/c\(^2\), and
(F)~\(P_{tot} = \left|\sum\vec{P}_i\right|<\) 250 MeV/c.
The criterion (A) corresponds to a loose energy cut which reduces the
background without much computation needed.  As stated above, it is
possible for one of the photons to be invisible, for which (B) allows.
If there are 3 rings criterion (C) requires that 2 of the rings
reconstruct to give a \(\pi^0\) mass.  Since there are no muons nor
charged pions expected, no decay electrons should be found and any
events which have them will be cut by (D).  Finally (E) requires the
total invariant mass to be near that of the proton and (F) requires
the total reconstructed momentum (magnitude of the vector sum of all
individual momenta) to be below the Fermi momentum for \(^{16}\)O.
Figure~\ref{fig:peppo-massmom} shows distributions of \peppo{} MC,
atmospheric MC, and data in reconstructed momentum {\it vs.}  invariant mass
after criteria (A)-(D) have been applied.  Criteria (E) and (F) are shown
by the box.

When these criteria are applied to 45 kton\(\cdot\)years (736 days) 
of data Superkamiokande
finds no candidate events.  Using atmospheric neutrino background MC
equivalent to 40 years of data taking it is estimated that 0.2
background events are expected in the data.  From MC simulations of
\peppo{} events, the efficiency to select any \peppo{} events in the
data sample is 44\%.  This gives a limit on the proton lifetime
divided by \peppo{} branching ratio (partial limit) of
\cllimit{\Mpeppo}{2.9}{33}.

\begin{table}[tbh]
\caption{\small Water: Analysis for the $p \rightarrow e^+   \;   \pi^0               $
 channel\protect\cite{Viren:1999pk}.}
{
\begin{tabular}{l|c|c|c}
 & \bf{$p \rightarrow e^+   \;   \pi^0               $} & MC & Data  \\
\hline
$6000<Q_{tot}<9500$ p.e. &  & & \\
2 or 3 e-like rings &  & & \\
$85 < M_{inv,\gamma\gamma} < 185$ MeV (if 3 rings) &  & & \\
no decay electron &  & & \\
$800 < M_{inv} < 1050$ MeV &  & & \\
 $\sum p<$ 0.250 GeV          &  44\% &   0.2 & 0  \\ 
\hline
\end{tabular}
\label{tab:1sk}
}
\end{table}

Superkamiokande searches for the \pnukp{} mode\cite{Viren:1999pk}
by looking for the products
from the two primary branches of the \(K^+\) decay: \(K^+ \rightarrow
\mu^+\nu_\mu\) and \(K^+ \rightarrow{} \pi^+\pi^0\).
In the \(K^+ \rightarrow
\mu^+\nu_\mu\) case, when the decaying proton is in the \(^{16}\)O,
the nucleus will be left as an excited \(^{15}\)N.  Upon
de-excitation, a prompt 6.3 MeV photon will be emitted.  So this
second branch has two independent searches: one in which the signature
of this prompt photon is required and one in which it is explicitly
absent.  

When searching for the \pnukpnumu{} with a 6.3 MeV prompt photon
search the following criteria are required:
(A)~1 \(\mu\)-like ring,
(B)~1 decay electron,
(C)~\(215 < P_{\mu} < 260\) MeV/c, and
(D)~\(N_{PMT}>7 ; 12 < t_{PMT} < 120\) ns before \(\mu\) signal.
The only particle giving a visible ring is the mono-energetic muon.
Criteria (A-C) select for that.  In criterion (D), \(t_{PMT}\) is the
time a PMT was hit subtracted by the time it would take a photon to
get directly from the fit event vertex to the PMT (so called ``time
minus time of flight'' or ``timing residual''). 

This search has an efficiency of 4.4\%, finds no candidates on an
estimated background of 0.4 events, and sets a limit of
\cllimit{\Mpnukpnumug}{2.1}{32}.

The complementary case where no prompt gamma is allowed has the same
criteria as the prompt gamma case except for the last:
(D)~\(N_{PMT}<=7 ; 12 < t_{PMT} < 120\) ns before \(\mu\) signal.
Since this allows a significant amount of background to survive the
selection criteria, the limit is set by fitting for an excess of proton
decay events above the atmospheric neutrino background in the
reconstructed momentum spectrum.  
In this region, 70 candidate events are found which is consistent with
the 74.5 events expected from atmospheric neutrinos.  With an
efficiency of 40\% a limit of \cllimit{\Mpnukpnumunog}{3.3}{32} is
found.

The criteria for the \pnukppippio{} search are as follows:
(A)~2 e-like rings,
(B)~1 decay electron,
(C)~\(85 < M_{inv,\pi^0} < 185\) MeV/c\(^2\),
(D)~\(175 < P_{\pi^0} < 250\) MeV/c,
(E)~\(40 < Q_{\pi^+} < 100\) PE.
The \(\pi^+\) is very close to Cerenkov threshold and is expected
to only produce a small amount of light as in (E).  Since this is not
enough to produce an identifiable ring only the rings from the 2
photons from the decay of the \(\pi^0\) are required in (A).  These
photons must reconstruct to an invariant mass in the range defined by
(C) as well as a momentum range defined in (D).
No candidates are found and 0.7 background events are expected.  The
selection efficiency is 6.5\%, giving a partial lifetime limit of
\cllimit{\Mpnukppippio}{3.1}{32}.

The combined limit for these three topologies is
\cllimit{\Mpnukpnumu}{6.8}{32}.

\subsection{Future plans}
Given the large existing experience in large underground Water Cerenkov detectors,
a certain number of conceptual ideas have been put forward in order to increase further
the mass by at least an order of magnitude compared to the largest Superkamiokande
detector. These are summarized in the last column of Table~\ref{tab:futurewater}. The size of these detectors
will eventually be limited by the light attenuation in water, however, this length is measured
to be in the range of 100~m in Superkamiokande. Hence, the limiting factor is rather driven
by the size of the underground cavern and by the cost of the experiment. A way to reduce
costs is to limit the effective photocathode coverage however at the cost of physics performance.
To set the scale, Superkamiokande has a design coverage of 40\%. A one-megaton HyperK\cite{Itow:2001ee} with similar
performance would contain between 100'000--200'000 large area (8 inch) photomultipliers.

\begin{table}[tbh]
\caption{Comparison of Water Cerenkov Detectors}
{
\begin{tabular}{@{}lccccccc@{}}
\hline
{} &{} &{} &{} &{}\\[-1.5ex]
Parameters & Kamioka & IMB-3 & SuperK & SNO & HyperK\cite{Itow:2001ee} & UNO\cite{Jung:1999jq} & 3M\cite{Diwan:2003uw}  \\[1ex]
& Japan & USA & Japan & Canada & (proposed) & (proposed) & (proposed) \\[1ex]
\hline
Mass kt & 4.5 & 8 & 50 & 8 & 1000 & 650 & 1000 \\[1ex]
Fiducial & 1.0 & 3.3 & 22 & 2 & 800 & 440 & 800 \\[1ex]
\hline
Effective & & & & & & & \\[1ex]
Photocathode & 20\% & 4\% & 40\% & 60\% & 20--40\% & $1/3$ 10\% & 14\%  \\[1ex]
coverage & & & & &  & $2/3$ 40\% &   \\[1ex]
\hline
\end{tabular}\label{tab:futurewater} }
\end{table}

\section{Large Liquid Argon imaging TPC}
\subsection{The technique}
Among the many ideas developed around the use of liquid noble gases, the Liquid 
Argon Time Projection Chamber (LAr TPC), conceived and
proposed at CERN by C.~Rubbia in 1977~\cite{intro1}, certainly represented one of the most
challenging and appealing designs.
The technology was proposed as a tool for
uniform and high accuracy imaging of massive detector volumes. 
The operating principle of the LAr TPC was based on
the fact that in highly purified LAr ionization tracks could indeed be transported
undistorted by a uniform electric field over distances of the 
order of meters~\cite{Aprile:1985xz}. Imaging is
provided by wire planes placed at the end of
the drift path, continuously sensing and recording the signals induced by
the drifting electrons. Liquid Argon is an ideal medium since it provides
high density, excellent properties (ionization, scintillation yields) and
is intrinsically safe and cheap, and readily available anywhere as a standard by-product
of the liquefaction of air.

The feasibility of this technology has been further
demonstrated by the extensive ICARUS R\&D program, which included
studies on small LAr volumes about proof of principle, LAr purification
methods, readout schemes and electronics, as well as studies with
several prototypes of increasing mass on purification technology,
collection of physics events, pattern recognition, long duration tests and
readout. The largest of these devices had a mass of 3 tons of
LAr~\cite{3tons,Cennini:ha} and has been continuously operated for more than four years, collecting
a large sample of cosmic-ray and gamma-source events. Furthermore, a smaller
device with 50 l of LAr~\cite{50lt} was exposed to the CERN neutrino
beam, demonstrating the high recognition capability of the technique for
neutrino interaction events.

The realization of the 600 ton ICARUS detector culminated with its full test 
carried out at surface during the summer 2001~\cite{t600paper}. This test demonstrated that
the LAr TPC technique can be operated at the kton scale with a drift length of 1.5~m.
Data taking of about 30000 cosmic-ray triggers have allowed to test
the detector performance in a quantitative way and results have been 
published in~\cite{Amoruso:2003sw,gg4,gg2,gg3,gg1}.

\subsection{A conceptual design for a 100~kton detector}
A conceptual design for a 100~kton LAr TPC was given in Ref.~\cite{Rubbia:2004tz}.
The basic design features of the detector can be summarized as follows:
(1) Single 100 kton ``boiling'' cryogenic tanker at atmospheric
pressure for a stable and safe equilibrium condition (temperature is constant while Argon is boiling).
The evaporation rate is small (less than $10^{-3}$ of the total volume per day given
by the very favorable area to volume ratio) and is compensated
by corresponding refilling of the evaporated Argon volume. 
(2) { Charge imaging, scintillation and Cerenkov light readout}
for a complete (redundant) event reconstruction. This represents a clear advantage over large
mass, alternative detectors operating with only one of these readout modes. The physics
benefit of the complementary charge, scintillation and Cerenkov readout
are being assessed.
(3)  { Charge amplification to allow for very long drift paths}. 
The detector is running in bi-phase mode. In order to allow for drift lengths as long as $\sim$ 20 m,
which provides an economical way to increase the volume of the detector with a constant number
of channels, charge attenuation will occur along the drift due to attachment to the remnant impurities present
in the LAr. This effect can be compensated with 
charge amplification near the anodes located in the gas phase.
(4)  { Absence of magnetic field}, although this possibility might be considered at a later
stage. Physics studies~\cite{Rubbia:2001pk} indicate that a magnetic field
is really only necessary when the detector is coupled to a Neutrino Factory.

The cryogenic features of the proposed design are based on the industrial
know-how in the storage of liquefied natural gases (LNG, $T\simeq 110$ K at 1 bar),
which developed quite dramatically in the last decades, driven by the petrochemical and space rocket industries. 
LNG are used when volume is an issue, in particular, for storage.
The technical problems associated to the design of large cryogenic tankers,
their construction and safe operation have already been addressed and engineering problems
have been solved by the petrochemical industry. 
The current state-of-the-art contemplates cryogenic tankers of
200000~m$^3$ and their number in the world is estimated to be $\sim$~2000
with volumes larger than 30000~m$^3$ with the vast majority built
during the last 40 years. 
Technodyne International Limited, UK~\cite{Technodyne}, which
has expertise in the design of LNG tankers, has been appointed to initiate a feasibility
study in order to understand and clarify the issues related to the operation of a large
underground LAr detector. A final report is expected soon.

\begin{figure}[htb]
\centerline{\epsfxsize=6in\epsfbox{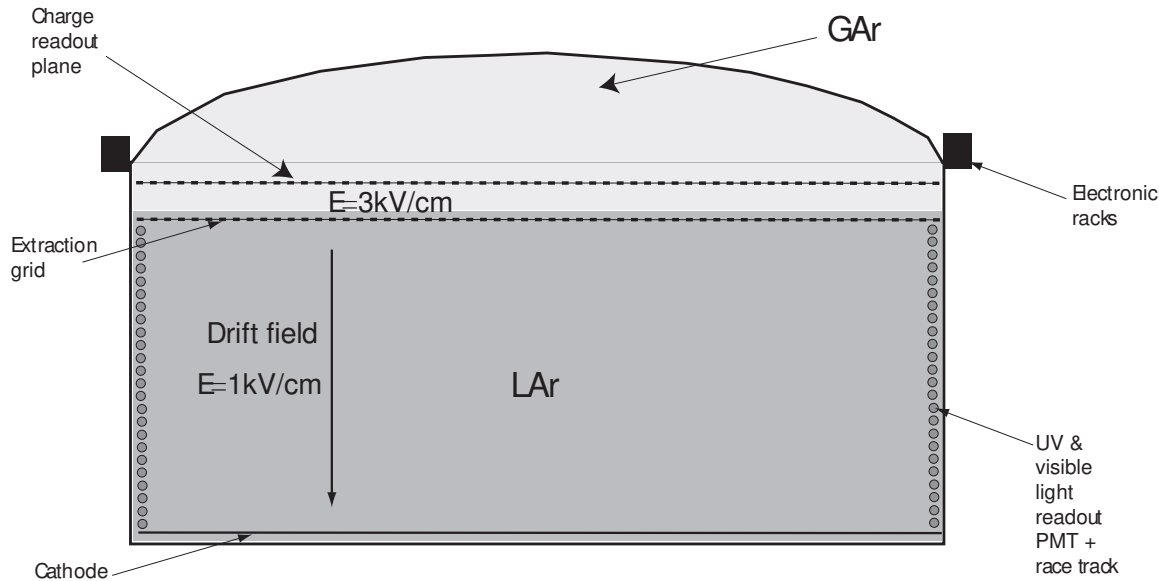}}   
\caption{Schematic layout of a 100 kton liquid Argon detector. The race track is composed of a set
of field shaping electrodes.}
\label{fig:t100schema}
\end{figure}

A schematic layout of the inner detector is shown in Figure~\ref{fig:t100schema}. The detector is characterized
by the large fiducial volume of LAr included in a large tanker, with external dimensions
of approximately 40 m in height and 70 m in diameter. A cathode located at the bottom of the 
inner tanker volume 
creates a drift electric field of the order of 1~kV/cm over a distance of about 20~m. 
In this field configuration ionization electrons
are moving upwards while ions are going downward. The electric field is delimited on the sides of the tanker
by a series of ring electrodes (race-tracks) placed at the appropriate potential by a voltage divider.

The tanker contains both liquid and gas Argon phases at equilibrium. Since purity is a concern for very long
drifts of 20 m, we assume that the inner detector could be operated in bi-phase mode:
drift electrons produced in the liquid phase are extracted from the liquid into the gas phase with
the help of a suitable electric field and then amplified near the anodes. 
In order to amplify the extracted charge one can consider various options: amplification
near thin readout wires, GEM~\cite{Sauli:qp} or LEM~\cite{Jeanneret:mr}. 
Studies that we are presently conducting show that gain factors of 100-1000 are achievable in pure Argon~\cite{dmrd}.
Amplification operates in proportional mode. Since the readout is limited to the top of the detector, 
it is practical to route cables out from
the top of the dewar where electronics crates can be located around the dewar outer edges.

\begin{figure}[ht]
\centerline{\epsfxsize=6.1in\epsfbox{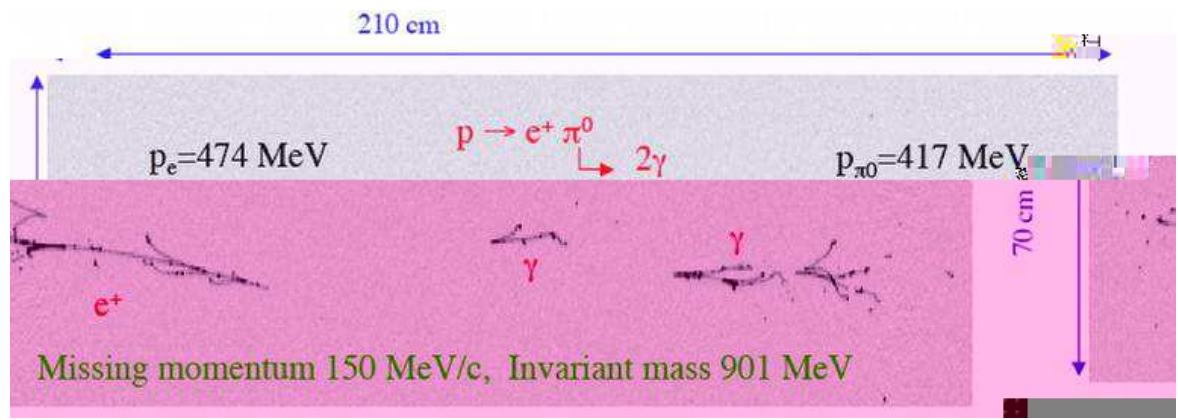}}   
\caption{MC event of $p\rightarrow e^+\pi^0$ in a liquid Argon TPC. \label{fig:larepi}}
\end{figure}

\begin{figure}[ht]
\centerline{\epsfxsize=6.1in\epsfbox{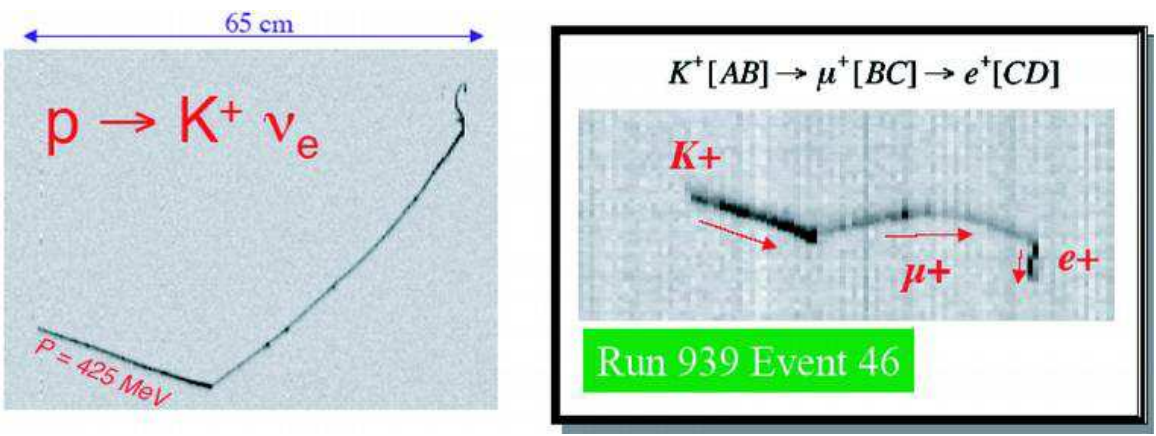}}   
\caption{(left) MC event of $p\rightarrow K^+\nu$ in a liquid Argon TPC. (right) Real event
collected in ICARUS T600 cosmic run performed on surface with a stopping kaon topology\label{fig:larnuk}}
\end{figure}

\begin{figure}[ht]
\centerline{\epsfxsize=4.1in\epsfbox{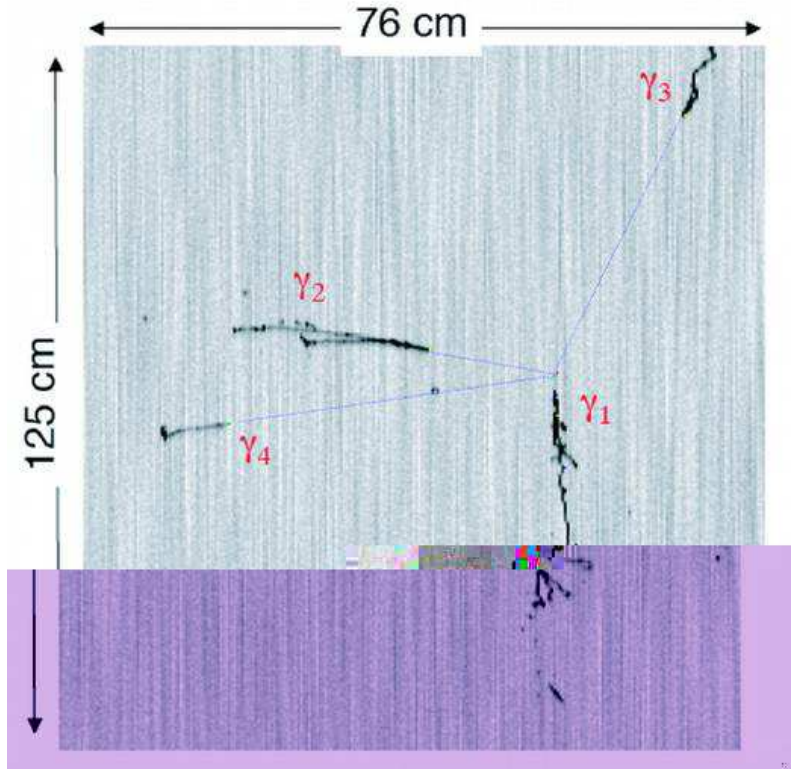}}   
\caption{MC event of $n\rightarrow \nu K^0\rightarrow \nu\pi^0\pi^0$ in a liquid Argon TPC. \label{fig:larnupipi}}
\end{figure}

After a drift of 20 m at 1 kV/cm, the electron cloud
diffusion reaches approximately a size of 3 mm, which corresponds to the
envisaged readout pitch. Therefore, 20 m practically
corresponds to the longest conceivable drift path. 
As mentioned above, drifting over such distances
will be possible allowing for some charge attenuation due to attachment
to impurities. If one assumes that the operating electron lifetime is at least $\tau\simeq 2$~ms (this is
the value obtained in ICARUS~T600 detector during the technical run~\cite{gg3} and
better values of up to $10$~ms were reached on smaller prototypes during longer runs),
one then expects an attenuation of a factor $\sim$ 150 over the distance of 20~m. 
This loss will be compensated by the proportional gain at the anodes.
We remind that the expected attenuation factor (compensated by the amplification) will not introduce
any detection inefficiency, given the value of $\sim$ 6000 ionization electrons per millimeter produced 
along a minimum ionizing track in LAr.

In addition to charge readout, one can envision to locate PMTs around the inner surface of the tanker. 
Scintillation and Cerenkov light can be readout essentially independently. 
LAr is a very good scintillator with about 50000 $\gamma$/MeV (at zero electric field). 
However, this light is essentially distributed around a line at $\lambda=128$~nm and, therefore, 
a PMT wavelength shifter (WLS) coating is required. 
Cerenkov light from penetrating muon tracks has been successfully detected in a LAr TPC~\cite{gg2};
this much weaker radiation (about $700\ \gamma/$MeV between 
160~nm and 600~nm for an ultrarelativistic muon) can be separately identified with
PMTs without WLS coating, since their efficiency for the DUV light
will be very small. 

A series of R\&D is ongoing to further develop the conceptual ideas outlined above and
are discussed in Ref.\cite{nuint04}.

\subsection{Estimated sensitivities to nucleon decay searches}
In absence of data, the sensitivity to nucleon decay has been studied carefully with 
MC simulations, which include all known physics and detector effects. Overall the searches
largely profit, regardless of the decay mode, of the clear imaging properties
which allows to unambiguously tag most events. In particular, low particle
detection thresholds are very effective at suppressing atmospheric neutrino
events. Examples of signal events are shown in Figures~\ref{fig:larepi},
\ref{fig:larnuk} and \ref{fig:larnupipi}. In Figure~\ref{fig:larnuk} we included
a real event collected in a ICARUS T600 cosmic run performed on surface.
The real event shows very well the topology of the kaon, muon decay chain.
In particular, the figure shows the power of multiple $dE/dx$ measurements.

For the $p\rightarrow e^+\pi^0$ channel, one finds that the $\pi^0$ is absorbed
via nuclear interaction about 45\% of the times. This is comparable to what
happens on Oxygen, as the probability for interaction scales with $A^{1/3}$, hence,
the ratio of rate of interaction
between Oxygen and Argon is $(A_{O}/A_{Ar})^{(1/3)}\approx 0.77$.

A list of cuts is presented in Table~\ref{tab:1lar}. Rates are normalized
to an exposure of 1000~kton$\times$year. The signal events are required to be balanced (up
to Fermi motion), with all particles identified as such and with a total invariant
mass compatible with that of a proton. After the requirement for one $\pi^0$
and one positron (or electron), only $\nu_e$ and $\bar\nu_e$ CC events survive.
It should be stressed that the study of the background did not at this stage
rely on a full simulation and reconstruction of events hence has to be considered
as preliminary. Work is in progress to better estimate reconstruction effects.
Finally a cut on reconstruction momentum and energy is sufficient to reject
all backgrounds, leaving less than 1 $\nu_e$ CC event. The signal
efficiency is 45\%.

\begin{table}[tbh]
\caption{\small Liquid Argon: Cuts for the $p \rightarrow e^+   \;   \pi^0               $
 channel. Survival fraction of 
 signal (first column) and background events through event 
 selections applied in succession.}
{
\begin{tabular}{c|c|c|c|c|c|c|c}\cline{2-8}
 & \bf{$p \rightarrow e^+   \;   \pi^0               $} &
\bf{$\nu_e$ CC}& \bf{$\bar{\nu}_e$ CC} & \bf{$\nu_{\mu}$ CC}&
\bf{$\bar{\nu}_{\mu}$ CC} & \bf{$\nu$ NC} & \bf{$\bar{\nu}$ NC} \\
\hline
                        &  100\% &  59861 &  11707 & 106884 &  27273 &  64705 &  29612 \\ \hline
One $\pi^0$                         &  54.0\% &  6604 &  2135 & 15259 &  5794 &  8095 &  3103 \\ \hline
One positron                        &  54.0\% &  6572 &  2125 &    20 &     0 &     0 &     0 \\ \hline
No charged pions                    &  53.9\% &  3605 &   847 &     5 &     0 &     0 &     0 \\ \hline
No protons                          &  50.9\% &  1188 &   656 &     1 &     0 &     0 &     0 \\ \hline
 $\sum p<$ 0.4 GeV          &  46.7\% &   454 &   127 &     0 &     0 &     0 &     0 \\ \hline
0.86 $<$ E $<$ 0.95 GeV   &  45.3\% &     $<1$ &     0 &     0 &     0 &     0 &     0 \\ \hline
\hline
\end{tabular}
\label{tab:1lar}
}
\end{table}

The $p\rightarrow \nu K^+$ channel profits from the expected 
excellent kaon identification capability. The probability that the kaon
interacts inside the nuclear matter is also very small since strangeness
should be conserved. A list of cuts is presented in Table~\ref{tab:1klar},
normalized to an exposure of 1000~kton$\times$year. At this level of sophistication
in the analysis, it appears that positive particle identification and a single
kinematical cut on the total energy is sufficient to totally reduce all backgrounds
keeping an efficiency of 97\% for the signal. This efficiency is about an order of magnitude
larger than that obtained in Water Cerenkov detectors, as illustrated in previous
sections. Work is in progress to further
understand these figures, in particular concerning strange production 
backgrounds due to cosmic ray muons as a function of the underground
depth\cite{mucrback}.
 
\begin{table}[tbh]
\caption{\small Liquid Argon: Cuts for the  $p \rightarrow K^+   \; \bar{\nu}$
 channel. Survival fraction of 
 signal (first column) and background events through event 
 selections applied in succession.}
{
\begin{tabular}{c|c|c|c|c|c|c|c}\cline{2-8}
\hline
\multicolumn{1}{c|}{\bf{Cuts}} & \bf{$p \rightarrow K^+   \; \bar{\nu}             $} &
\bf{$\nu_e$ CC}& \bf{$\bar{\nu}_e$ CC} & \bf{$\nu_{\mu}$ CC}&
\bf{$\bar{\nu}_{\mu}$ CC} & \bf{$\nu$ NC} & \bf{$\bar{\nu}$ NC} \\
\hline
                        &  100\% &  59861 &  11707 & 106884 &  27273 &  64705 &  29612 \\ \hline
One Kaon                            &  96.75\% &   308 &    36 &   871 &   146 &   282 &    77 \\ \hline
No $\pi^0$                          &  96.75\% &   143 &    14 &   404 &    56 &   138 &    25 \\ \hline
No positrons                        &  96.75\% &     0 &     0 &   400 &    56 &   138 &    25 \\ \hline
No muons                            &  96.75\% &     0 &     0 &     0 &     0 &   138 &    25 \\ \hline
No charged pions                    &  96.75\% &     0 &     0 &     0 &     0 &    57 &     9 \\ \hline
Total Energy $<$ 0.8 GeV            &  96.75\% &     0 &     0 &     0 &     0 &     $<1$ &     0 \\ \hline
\hline
\hline
\end{tabular}
\label{tab:1klar}
}
\end{table}

\section{Large Water Cerenkov or Liquid Argon TPC ? or both?}

\subsection{One versus the other?}
Among all the technologies considered today, 
the Water Cerenkov and the liquid Argon TPC are the only ones that have the highest
potential to provide a rich physics program, 
including both accelerator and non-accelerator aspects. Therefore, it is a relevant
question to ask which detector should be adopted for the next generation massive underground
detector on the timescale of the next decade. On one hand, the liquid Argon TPC
is a superior imaging technology than the Water technology. On the other hand, the experience
in massive underground Water Cerenkov detector is much greater than in the case of the liquid Argon TPC.

Indeed, the Water Cerenkov technology, is a very robust, proven and well-understood technology.
It was demonstrated to the 50 kton scale with Superkamiokande, which was operated underground
for many years. A megaton-scale detector,
like the HyperKamiokande one, is perceived as a  ``straight-forward'' extrapolation 
of Superkamiokande. Provided that a megaton-scale detector is 
financeable, the question that needs to be understood is whether a twenty-fold increase
in statistics compared to Superkamiokande will answer the important physics questions of the
next decade.

On the other hand, the liquid Argon TPC is a new and challenging technology.
It is the fruit of many years of R\&D effort conducted by the ICARUS collaboration. 
It was demonstrated up to the 0.6 kton scale on the surface test in summer 2001.
An extrapolation from 0.6 to 100 kton is certainly a big step. A conceptual design based on the well-proven LNG 
technology was proposed and is being elaborated. Nonetheless, 
extrapolation to 100~kton might require an intermediate step,  a 10\% prototype ($\simeq$ 10 kton),
to prove that the LNG technology and the physics can be fruitfully married in an actual experiment.
In addition to the mass increase, the imaging provided by such a detector would provide a qualitative and 
quantitative improvement in the physics program.
R\&D efforts are on-going and must be vigorously pursued.
In parallel, high statistics, precision physics will require a 
$\simeq$100~ton detector acting as a near experiment in a neutrino beam. This step
is mandatory in order to improve the knowledge of neutrino interactions on Argon and the
ability to reconstruct them with a liquid Argon TPC.

\subsection{Both -- complementarity}
\begin{figure}[ht]
\centerline{\epsfxsize=3in\epsfbox{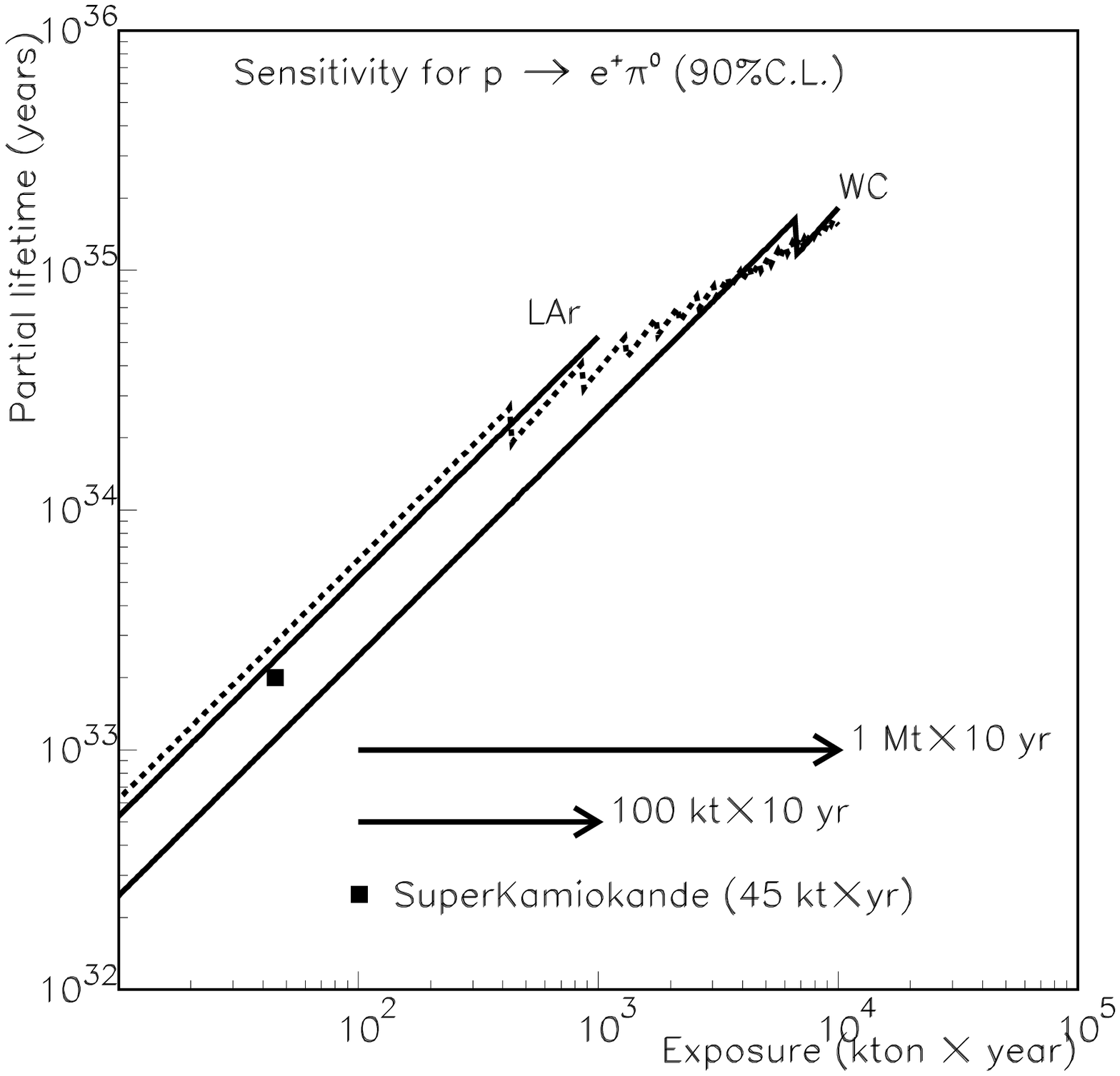}\epsfxsize=3in\epsfbox{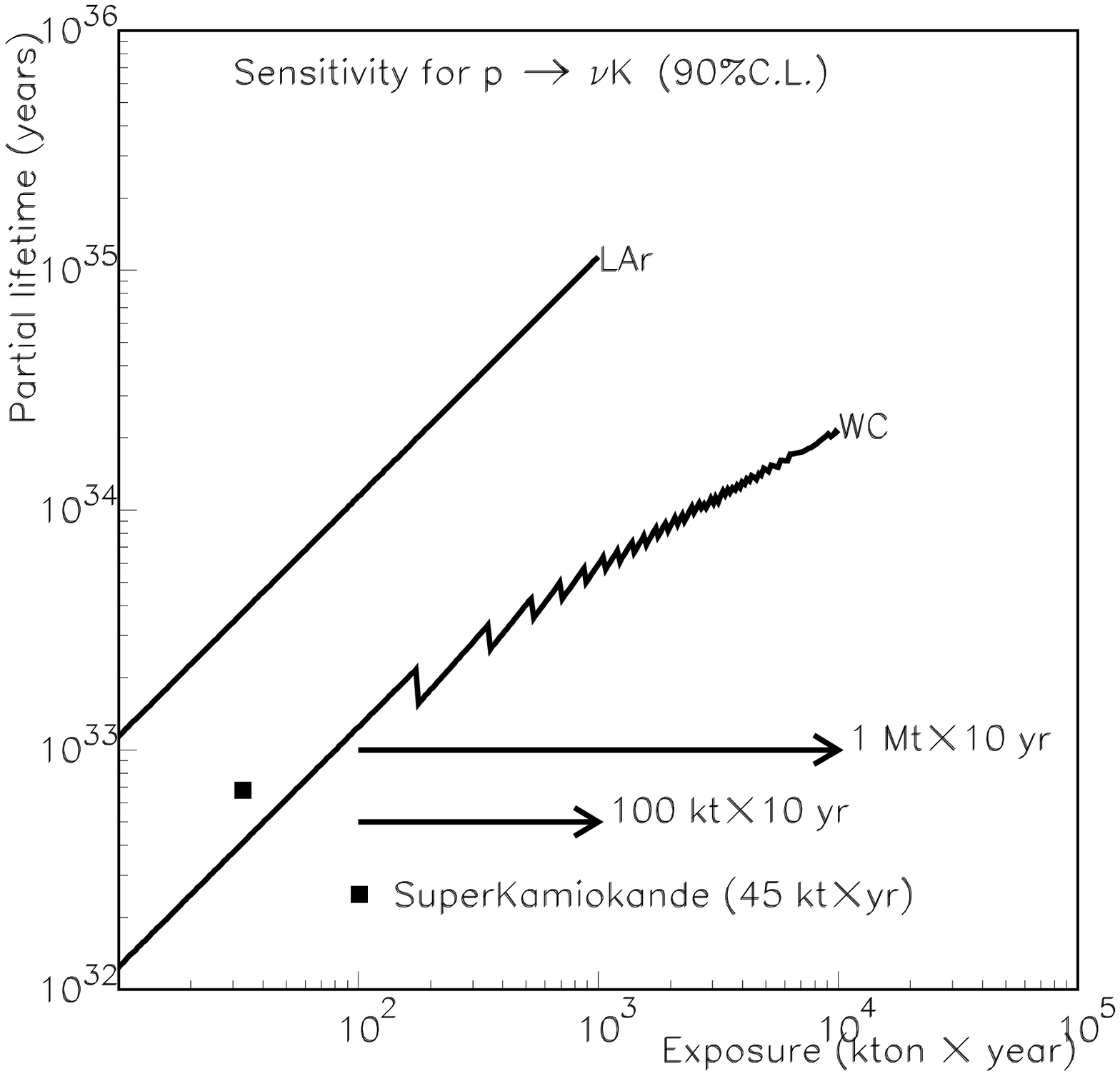}}   
\caption{Expected sensitivity of Water Cerenkov (WC) and liquid Argon TPC (LAr) detectors
for $p\rightarrow e^+\pi^0$ and $p\rightarrow \nu K$ decays. For WC, we assumed similar 
performance as in SuperK. The dashed line for the $p\rightarrow e^+\pi^0$  channel corresponds
to the standard SK cut, while the thick line corresponds to the ``tight'' cut analysis.}
\label{fig:compsensi}
\end{figure}

In an ideal configuration, an optimum physics program would exploit the complementarity
between the two techniques in order to best answer the physics questions. We illustrate
this complementarity in the case of the proton decay searches in Figure~\ref{fig:compsensi},
where the sensitivity to $p\rightarrow e^+\pi^0$ and $p\rightarrow \nu K^+$ are plotted
as a function of the exposure for the two techniques (WC and LAr). In this plot, the contribution
of the background has been included. This explains the see-saw behavior of the sensitivities
as soon as backgrounds are greater than 1 event. In this case, we always assumed that the
actually observed number of events is identical to the expected (to the nearest integer of course).
The two graphs show that: (1) for $p\rightarrow e^+\pi^0$ the larger exposure of WC is important;
(2) for the $p\rightarrow \nu K$ mode, the imaging of LAr largely compensates for the reduced mass.
This kind of argument can be extended to other decay modes, by realizing that WC detectors are most
sensitive to electromagnetic and back-to-back configurations, but are rather limited in tracking
performance, in detection of heavy or slow charged particles, and in multi-particle final-states. 

Table~\ref{tab:compwaterliqar} summarizes these features in a more quantitative way.
We also list in the Table relevant figures for other non-accelerator physics topics, like
the observation of supernova explosions (SN), and solar and atmospheric neutrinos.
Once again the complementarity between the two techniques, assuming a 650~kton water
Cerenkov and a 100~kton liquid Argon TPC, is visible.

\begin{table}[tbh]
\caption{Comparison of Water Cerenkov and Liquid Argon TPC detector}
{
\begin{tabular}{@{}lcc@{}}
\hline
& Water Cerenkov (UNO) & Liquid Argon TPC  \\[1ex]
\hline
Total mass & 650 kton & 100 kton  \\[1ex]
\hline
$p\rightarrow e\pi^0$ in 10 yrs &  $1.6\times 10^{35}$ yrs & $0.5\times 10^{35}$ yrs  \\[1ex]
& ($\epsilon\approx 17\%$, $\approx$ 1 evts background) &   ($\epsilon\approx 45\%$, $\approx$ 1 evt background)\\[1ex]
$p\rightarrow \nu K$ in 10 yrs & $0.2\times 10^{35}$ yrs & $1.1\times 10^{35}$ yrs  \\[1ex]
& ($\epsilon\approx 8.6\%$, $\approx$ 57 evts background) &   ($\epsilon\approx 97\%$, $\approx$ 1 evt background)\\[1ex]
$p\rightarrow \mu\pi K$ in 10 yrs & N/A & $8\times 10^{34}$ yrs  \\[1ex]
&  &   ($\epsilon\approx 98\%$, $\approx$ 1 evt background)\\[1ex]
\hline
SN cool off \@ 10 kpc & 194000 & 38500 (all flavors) \\[1ex]
& (mostly $\bar\nu_e p \rightarrow e^+ n$) & (64000 if nh-L mixing) \\[1ex]
SN in Andromeda & 40 events & 7 \\[1ex]
& & (12 if nh-L mixing) \\[1ex]
SN burst \@ 10 kpc & $\approx$ 330 $\nu-e$ elastic scattering & 380 $\nu_e$ CC (flavor sensitive) \\[1ex]
SN relic & Yes & Yes \\[1ex]
\hline
Atmospheric neutrinos & 60000 events/yr & 10000 events/yr \\[1ex]
\hline
Solar neutrinos & $E_e> 7 MeV$ & 324000 events/yr \\[1ex]
& (central module) & with $E_e> 5 MeV$\\[1ex]
\hline
\end{tabular}\label{tab:compwaterliqar} }
\end{table}

\section{Conclusion}
Massive tracking calorimeters were developed as ideal tools to look for nucleon
decays and for a wide underground physics program.
The increase in mass of the fine grain tracking calorimeters (NUSEX, Fr\'ejus and SOUDAN)
became quickly limited. The main scaling problem
of these technologies is that channel count (and therefore cost)  scale with the mass of the detector, since
they instrument the volume. Hence, detectors in the tens of kton range with high
granularity are impractical.

Water Cerenkov (WC) detectors do not possess the fine granularity of the ``tracking calorimeters'',
however, can be built in very large sizes since the instrumentation scales with the surface
of the detector and not the volume.
WC detectors provided very stringent {\it limits} on nucleon decays, thanks to their
mass scales. Given their coarse imaging, their sensitivity is limited to mostly electromagnetic
back-to-back nucleon decay events and given the background levels in many channels,
their sensitivity is expected
to scale with the square root of the their mass and exposure. The era of the linear
sensitivity gain with mass is over for various channels.

In the meantime, the liquid Argon imaging TPC has reached a high level of maturity thanks
to the extensive R\&D conducted by the ICARUS Collaboration during many years.
Today, quantitative progress in nucleon decay is calling for an application of this
technology at the 100~kton mass scale. We have shown a new conceptual design for such
a detector based on the LNG technology.

The megaton Water Cerenkov detector represents the conservative approach, while
the ``100 kton liquid Argon'' is certainly a very challenging design.
Given the foreseeable timescale of more than a decade for the 
next generation of massive underground detectors, it is our conviction that
a certain level of risk and challenge represents the most attractive
way towards potential progress in the field.

\section*{Acknowledgments}
We thank I.~Gil-Botella, A.~Ereditato, A.~Meregaglia and M.~Messina
for their help in the preparation of this presentation.
This work was supported by ETH/Zurich and the Swiss National Research Foundation.

\end{document}